\documentclass[prc,amsmath,amssymb,twocolumn,floatfix, nofootinbib,showpacs]{revtex4-1}

\usepackage{graphicx,bm}
\usepackage[tight]{subfigure}
\usepackage{color}

\newcommand{\bi}{\begin{itemize}}
\newcommand{\ei}{\end{itemize}}

\newcommand{\beq}{\begin{equation}}
\newcommand{\eeq}{\end{equation}}
\newcommand{\bea}{\begin{eqnarray}}
\newcommand{\eea}{\end{eqnarray}}
\newcommand{\ba}{\begin{align}}
\newcommand{\ea}{\end{align}}

\newcommand{\fmi}{\, \text{fm}^{-1}}

\newcommand{\mystrut}{\rule[-1mm]{0mm}{5.5mm}}

\newcommand{\Hdiag}{H_{\rm diag}}
\newcommand{\Trel}{T}
\newcommand{\GsExp}{G_s^{\rm exp}}
\newcommand{\GsInv}{G_s^{\rm inv}}
\newcommand{\Trelfull}{T_{\rm rel}}

\newcommand{\kcut}{k_{\rm cut}}
\newcommand{\lambdaequiv}{\lambda_{\rm eq}}
\usepackage[normalem]{ulem}
\usepackage{amssymb}

\begin{document}

\title{The Similarity Renormalization Group with Novel Generators}

\author{W. Li}\email{li.1287@osu.edu}
\affiliation{Department of Physics, The Ohio State University, Columbus, OH 43210}
\author{E.R.\ Anderson} \email{anderson@physics.ohio-state.edu}
\affiliation{Department of Physics, The Ohio State University, Columbus, OH 43210}
\author{R.J.\ Furnstahl}
\email{furnstahl.1@osu.edu}
\affiliation{Department of Physics, The Ohio State University, Columbus, OH 43210}

\date{\today}

\begin{abstract}
The choice of generator in the Similarity 
Renormalization Group (SRG) flow equation determines the evolution
pattern of the Hamiltonian. 
The kinetic energy has been used in the generator for 
most prior applications to nuclear
interactions, and other options have been largely unexplored.  
Here we show how variations of this standard choice can allow the
evolution to proceed more efficiently without losing its advantages.
\end{abstract}

\smallskip

\maketitle


\section{Introduction}
\label{sec:intro}

\begin{figure}[tb]
  \includegraphics[width=.9\columnwidth]{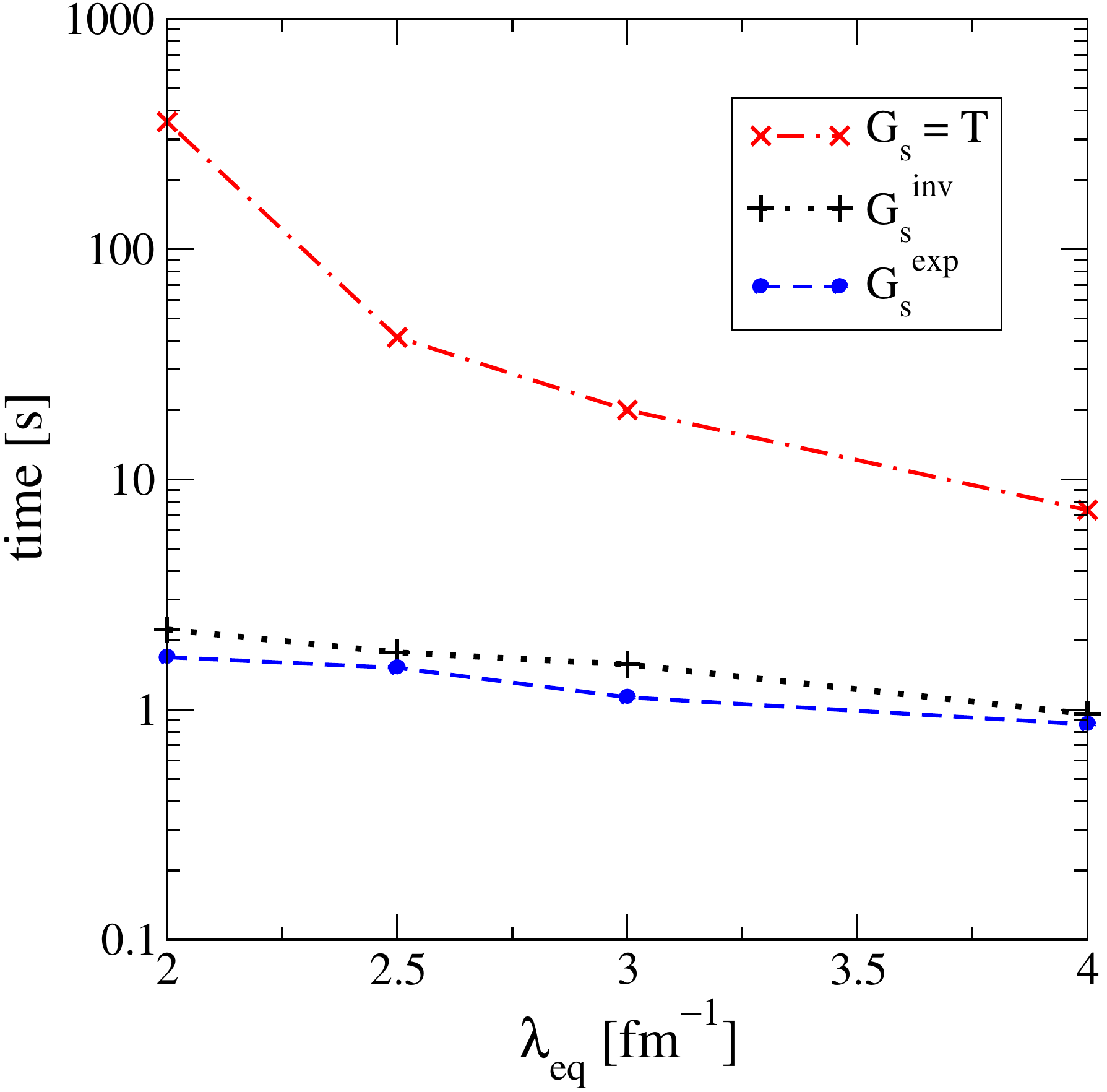}
  \vspace*{-.1in}
  \caption{(Color online) Computational time to evolve the Argonne $v_{18}$ $^1$S$_0$ 
  potential~\cite{Wiringa:1994wb}
  as a function of the final value of $\lambdaequiv$ (see text) for
  several generators, with $\sigma = 2\,\fmi$ for 
  $G_s^{\rm inv}$ and $G_s^{\rm exp}$.}
  \label{fig:time1}
\end{figure}

The Similarity Renormalization Group (SRG) uses a continuous
series of unitary
transformations to decouple high-momentum and low-momentum physics
in an input Hamiltonian~\cite{Glazek:1993rc,Wegner:1994}.
This decoupling means that expansions of physical observables 
generally become more convergent.
The SRG can be implemented through a flow equation for the evolving
Hamiltonian $H_s$,
\beq
  \frac{dH_s}{ds} = [ \eta_s, H_s ] = [ [ G_s, H_s ], H_s ]
  \;,
  \label{eq:srgflow}
\eeq
where $s$ is a flow parameter~\cite{Wegner:1994,Kehrein:2006} and
the generator $\eta_s$ is specified by the operator $G_s$.
With $G_s$ chosen to be
the relative kinetic energy $\Trelfull$, the SRG
has been applied successfully over the past few years to calculate
nuclear structure and reactions~\cite{Bogner:2006pc,Bogner:2007rx,Jurgenson:2009qs,Hebeler:2010xb,Bogner:2009bt,
Navratil:2010jn,Navratil:2010ey,Jurgenson:2010wy,Hergert:2011eh}.  However, different choices for $G_s$ will give rise to different  patterns of evolution, which may be advantageous.  
In this paper, two alternatives to $\Trelfull$ are evaluated for their effectiveness in decoupling and, in particular, for improvements in computing speed (see Fig.~\ref{fig:time1} for a representative example).
Our tests are for realistic nucleon-nucleon (NN) interactions in two-body
systems and for a one-dimensional model Hamiltonian applied to 
few-body bound states.

We focus on novel generators that have $G_s$ as functions of $\Trelfull$.  (Note: we can just as well consider the full kinetic energy
$T$ in our discussion, because the center-of-mass part commutes with the 
running Hamiltonian $H_s$, so we will use $T$ for convenience.)
In particular, we explore the ``inverse'' $G_s$ operator 
\beq
G_s = -\frac{\sigma^2}{1 + \Trel/\sigma^2} \equiv G_s^{\rm inv}
  \;,
  \label{eq:invgen}
\eeq
and the ``exponential'' $G_s$ given by
\beq
 G_s = -\sigma^2 e^{-\Trel/\sigma^2} \equiv G_s^{\rm exp}
 \;.
   \label{eq:expgen}
\eeq
Each has a Taylor series that  reduces to $\Trel$ (up to a constant,
which drops out from the commutator)
at low momentum or when \(\sigma\) is large.  As such, the independent parameter $\sigma$ controls the
separation of a low-energy region where $G_s$ behaves as
$T$ and the potential is driven toward the diagonal, and a high-energy region where evolution is suppressed.   This suppression can result in a significant computational speedup of the
SRG evolution when compared to calculations with \(G_{s}=\Trel\), while not impacting the advantageous properties of the evolution for low-momentum
applications.

The generators' suppression of running in unneeded parts of the Hamiltonian could mitigate the difficulties
of evolving  very large matrices for calculations of light 
atomic nuclei~\cite{Jurgenson:2010wy}, opening the door to more
tailored SRG generators and more effective evolution of three- and eventually
four-body interactions.
Even at the two-body level there are problems when trying to evolve
to large values of the flow parameter $s$. 
A recent example is a study of SRG decoupling with large-cutoff effective
field theory (EFT) potentials, which require evolution beyond the range usually
considered~\cite{Wendt:2011qj}.  In doing so, the SRG differential equations can become extremely stiff and take a prohibitively long time (weeks on a single processor) to evolve. 
This problem has hindered exploratory studies 
into issues such as what happens when a chiral EFT is
evolved to the regime of pionless EFT.

In Sec.~\ref{sec:NN}, 
we give representative results for applications to two-body
systems, including an analysis of the flow pattern.
These results are extended to few-body systems in Sec.~\ref{sec:oneD}
using a model one-dimensional Hamiltonian that has proved useful
in past applications~\cite{Jurgenson:2008jp,Anderson:2010aq}.
We summarize and outline future studies
in Sec.~\ref{sec:summary}.


\section{Two-nucleon systems}\label{sec:NN}

In this section we give representative results for evolving
realistic nucleon-nucleon potentials
using the novel generators from Eqs.~\eqref{eq:invgen} and \eqref{eq:expgen} 
in comparison to the usual
choice of $G_s = \Trel$.

\subsection{Performance}

The key advantage of the generators that we highlight is the improvement in computational performance.  For example, the time needed to evolve the Argonne $v_{18}$ $^1$S$_0$ potential~\cite{Wiringa:1994wb} to equivalent levels of decoupling with several generators is plotted in Fig.~\ref{fig:time1}.    The parameter $\lambda \equiv 1/s^{1/4}$, which has dimensions of a
momentum,  has been used to identify the momentum decoupling scale.
However,
different generators will evolve a given potential at different rates, so
comparing results with the same definition of $\lambda$ can be misleading. 
Therefore, we identify an ``equivalent'' $\lambdaequiv$ for each generator
that equalizes the degree of decoupling compared to $G_s = T$ (for which \(\lambdaequiv=\lambda\) by definition); the details are described 
in Sec.~\ref{subsec:decoupling}.  

The value of \(\sigma\) will also have an impact, as discussed below; in Fig.~\ref{fig:time1} we use the intermediate value \(\sigma=2\fmi\). With this choice, there is nearly an order of magnitude difference in the time to evolve the Argonne
$v_{18}$ potential with  \(\GsExp\) and \(\GsInv\) compared to \(G_s=\Trel\) at \(\lambdaequiv=4\fmi\) and two orders of magnitude by \(\lambdaequiv=2\fmi\). 
Note that nuclear interactions typically have been evolved for nuclear structure studies in the range $\lambda=1.5$--$2.2\fmi$.
The speed gains will depend on the initial potential and can be 
much less for the evolution of softer initial potentials; e.g., for the
N$^3$LO 500\,MeV chiral EFT potential of Ref.~\cite{Entem:2003ft}, evolving
with $\GsExp$ to $\lambdaequiv = 2\fmi$ is about 1.5 times as fast as 
with $G_s = \Trel$ and about 3 times as fast to $\lambdaequiv = 1.5\fmi$.

The numerical solution of the SRG evolution equations requires repeated dense
matrix-matrix multiplications.  As a consequence, the evolution has been carried
out on shared-memory computer architectures.  Recent calculations  using the SRG
with many-body forces are approaching the limits of what is practical  to evolve
in  memory on a single node because of the size of the model space needed 
\cite{Jurgenson:2010wy}.  A distributed scheme to solve the equations  would
permit larger model spaces to be utilized; however, the dense matrix
multiplication would then be limited by internode communication times.   The
reduced number of operations required by novel generators might help make
such a scheme possible.

\begin{figure}[b]
 \includegraphics[width=.9\columnwidth]{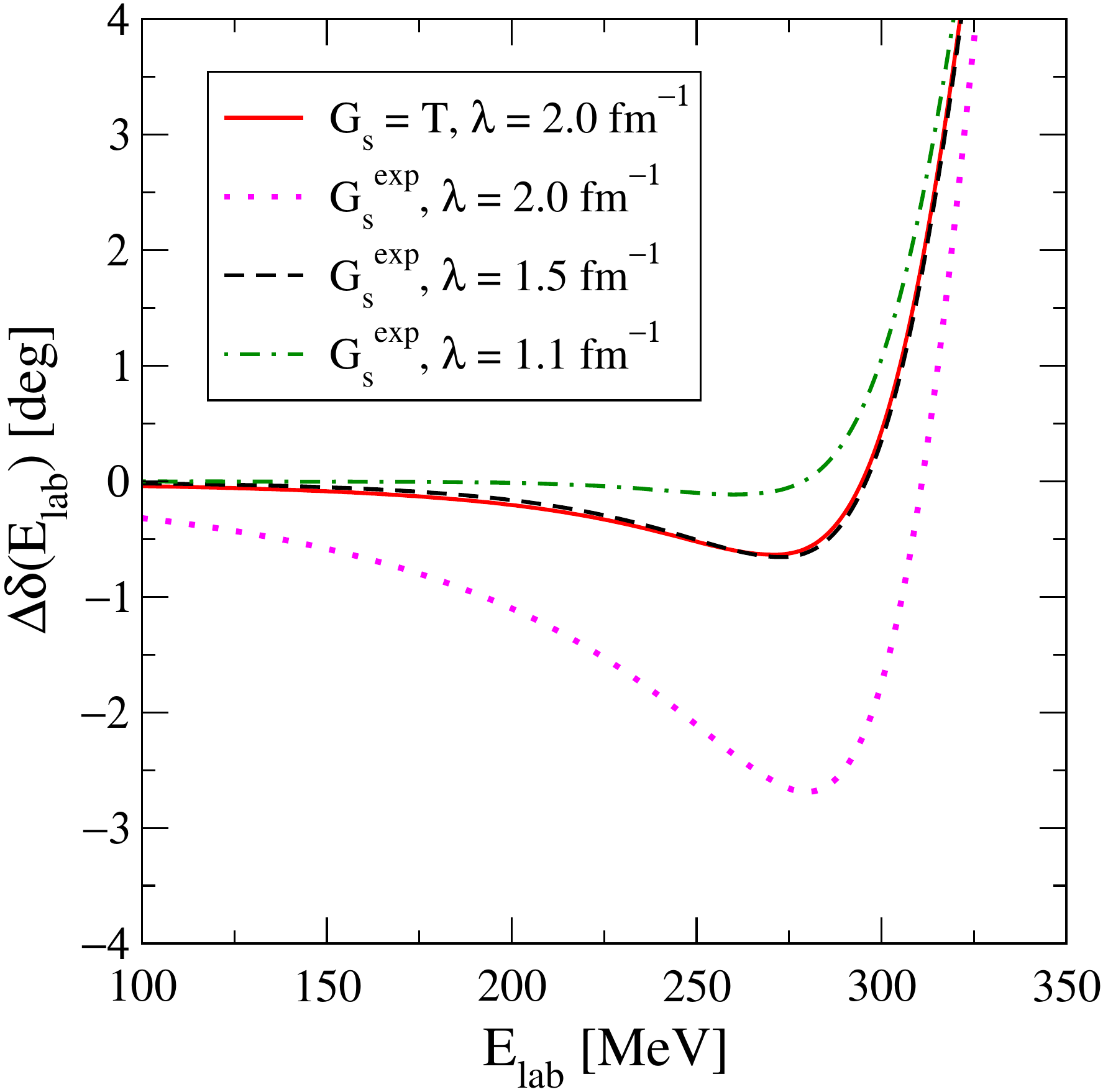}
 \caption{(Color online) Deviation of phase shifts calculated with the Argonne $v_{18}$ $^1$S$_0$ potential~\cite{Wiringa:1994wb} evolved
 with $G_s = T$ and  $G_s^{\rm exp}$ ($\sigma = 2\,\fmi$) to various
 $\lambda$ values and then 
truncated at $\kcut = 2\,\fmi$ to test decoupling. 
Phase shifts from untruncated potentials agree precisely with those from the initial potential.}
 \label{fig:phaseequiv}
\end{figure}

\begin{figure*}[tbh!]
   \subfigure[]{\includegraphics[width=0.7\textwidth]{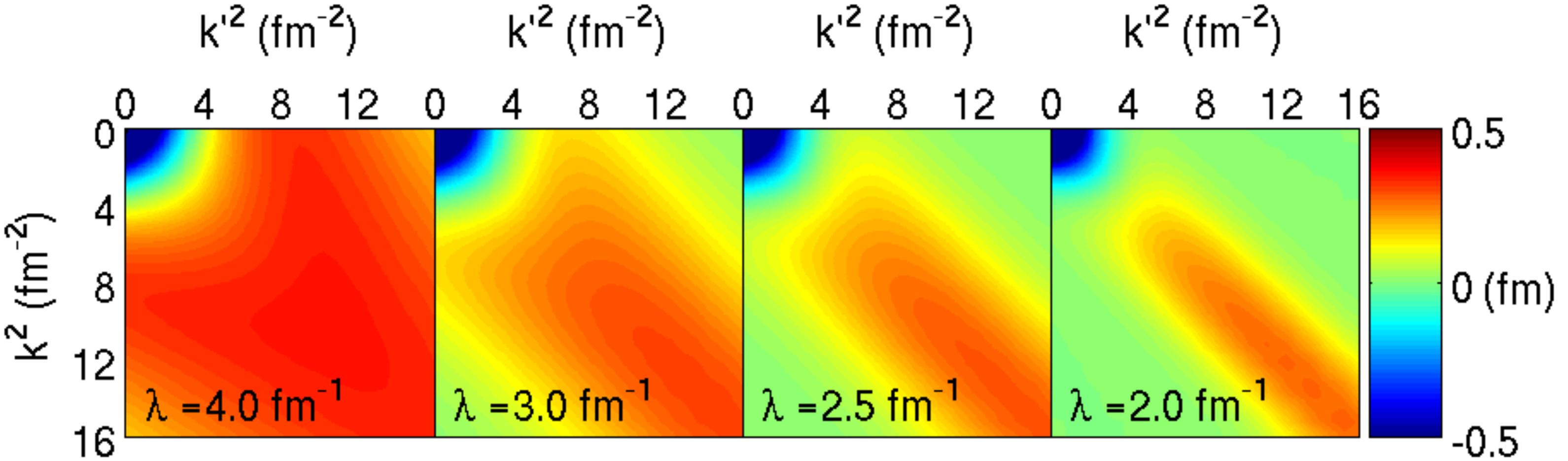}}
  \subfigure[]{\includegraphics[width=0.7\textwidth]{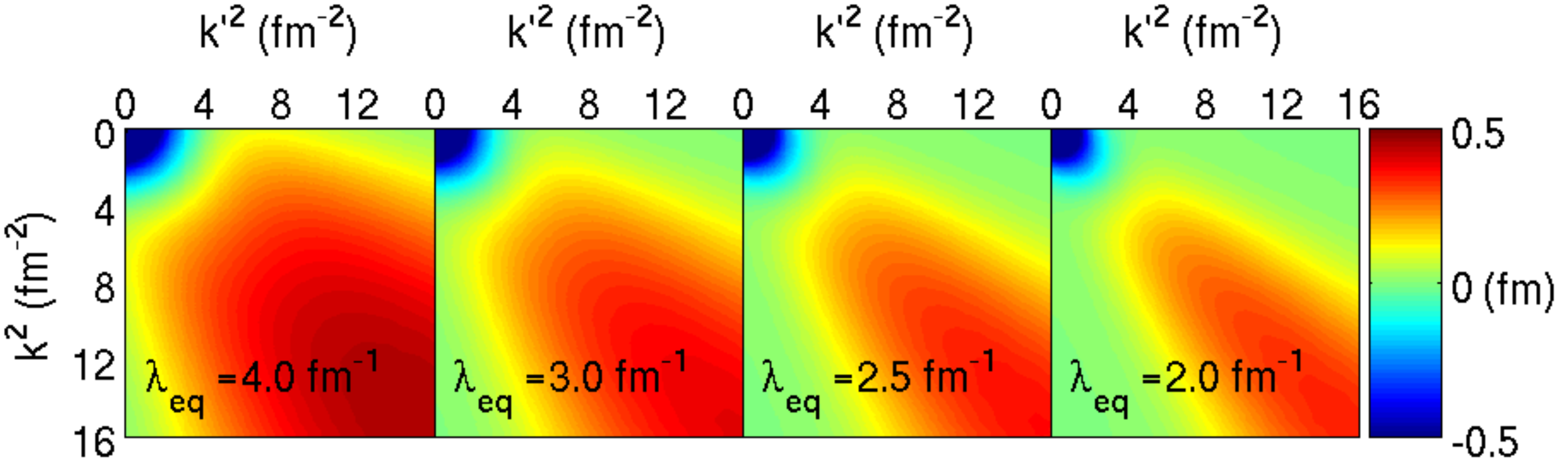}}
  \subfigure[]{\includegraphics[width=0.7\textwidth]{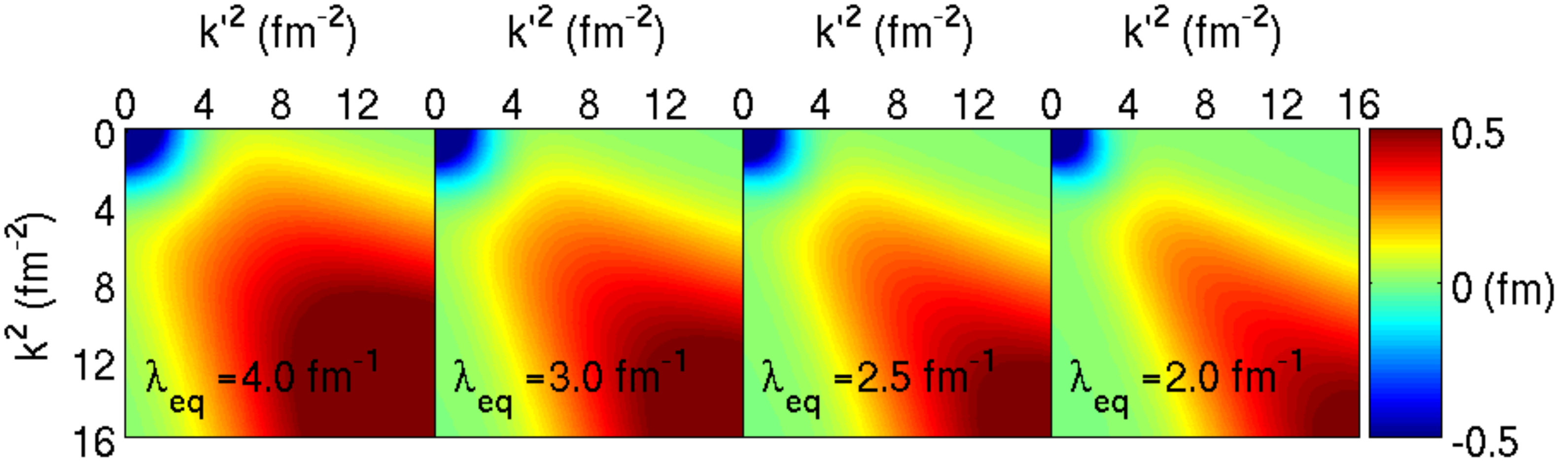}}
  \vspace*{-.1in}
  \caption{(Color online) Contour plots showing the evolution via
  Eq.~\eqref{eq:srgflow} at several values of $\lambdaequiv $
  starting from the momentum-space Argonne $v_{18}$ $^1$S$_0$ potential~\cite{Wiringa:1994wb}
  using  (a) $G_s = \Trel$, (b) $G_s^{\rm inv}$ from Eq.~\eqref{eq:invgen} with $\sigma = 2\,\fmi$, and (c) $G_s^{\rm exp}$ from Eq.~\eqref{eq:expgen} with $\sigma = 2\,\fmi$.\\ }
  \label{fig:evolveAll}
\end{figure*}

\subsection{Decoupling and $\lambdaequiv$}\label{subsec:decoupling}

To validate the apparent computational advantages of these generators, one must confirm that the decoupling characteristics of the  $G_s = T$  generator are also reproduced, so that  calculations of physical observables also become more convergent. However, if we evolve to the same $\lambda$, the
degree of decoupling for identical initial potentials differs  
for  \(\GsExp\)  and \(\GsInv\) compared to   $G_s = T$.   
These differences are evident in the deviations of $^1$S$_0$
phase shifts calculated from the evolved potentials using   \(\GsExp\)  and  $G_s = T$  from the unevolved potential shown in Fig.~\ref{fig:phaseequiv} for several different \(\lambda\) values (only $\GsExp$ is shown; \(\GsInv\) behaves similarly). 
If the full potentials were used, the phase shifts would agree -- up to numerical precision -- with those from the initial potential, because the evolution
in all cases is unitary.
However, the degree of decoupling for a given value of $\lambda$
can be made manifest by first cutting off
the potential  (that is, setting its matrix elements to zero)
above some value of $k$ and then calculating the phase shifts.  
In Fig.~\ref{fig:phaseequiv}, for
illustration, we choose the cutoff value $\kcut$ to be $2\fmi$.  
The signature of decoupling is that the phase shifts agree at lower energies 
and only deviate close to and above the cutoff.  
This is typically observed when the potential is evolved
so that $\lambda$ is less than the cut momentum~\cite{Jurgenson:2007td,Bogner:2007jb}.

This behavior provides us with a pragmatic way to define $\lambdaequiv$: identifying it with the decoupling behavior that is found for $G_s = T$
for a given $\lambda$. 
We use the results in Fig.~\ref{fig:phaseequiv} to illustrate the
procedure.
The continuous curve shows the deviation of phase shifts for the potential evolved with $G_s = T$ to $\lambda = 2\,\fmi$ and cut at $k = 2\,\fmi$.
As expected from decoupling, the deviation is small 
up to roughly $\kcut$.
We use this level of agreement as the criterion for identifying equivalent $\lambda$'s for other generators. That is, a potential is evolved to a series of $\lambda$ values with a different $G_s$ and then cut at $\kcut$ after evolution. The phase shifts are then compared to the level of decoupling observed for  $G_s = T$ at  $\kcut = \lambda$.  The approximate point in the novel $G_s$ evolution for which the cut phase shifts agree is equated with $\lambdaequiv$.
Consider the phase shifts for the potential evolved with $G_s^{\rm exp}$ to several values of $\lambda$ and cut at $k = 2\,\fmi$, as shown in Fig.~\ref{fig:phaseequiv}. For $G_s^{\rm exp}$, $\lambda$ evolved to $1.5\,\fmi$ gives a similar degree of decoupling as $\lambda$ evolved to $2\,\fmi$ with $G_s = T$.  As a result, we define $\lambda = 1.5\,\fmi$ to be the $\lambdaequiv = 2\,\fmi$ for $G_s^{\rm exp}$.
We use $\lambdaequiv$ for most comparisons in the following discussion.

\subsection{Flow Analysis}

Having chosen a working definition for the decoupling scale of evolution with  our novel generators, we take a closer look at the properties of the evolved potentials. In particular, we would like to see how a potential flows with evolution using these generators compared to \(G_s=\Trel\) and to understand how the choice of \(\sigma\) affects this flow. 

\begin{figure*}[tbh]
 \includegraphics[width=0.75\textwidth]{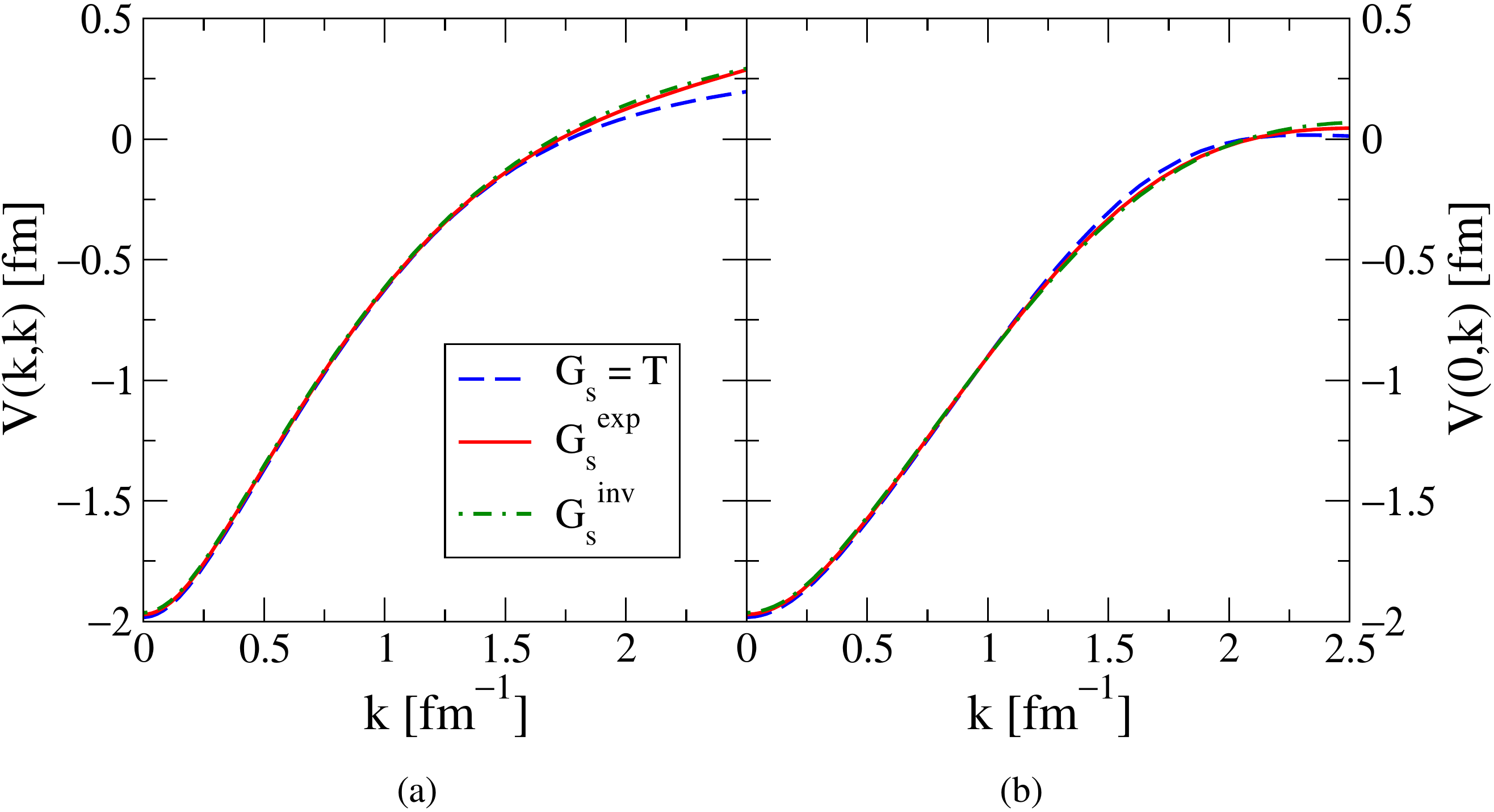}
 \caption{(Color online) Diagonal (a) and off-diagonal (b) momentum-space matrix elements for evolved 
 Argonne $v_{18}$ $^1$S$_0$ potential~\cite{Wiringa:1994wb} with $G_s = T$ and generators $G_s^{\rm exp}$ and $G_s^{\rm inv}$ at  \(\lambda=2\fmi\) with $\sigma = 2\,\fmi.$}
 \label{fig:universalitygenerator}
\end{figure*}

\begin{figure*}[tbh!]
 \includegraphics[width=0.75\textwidth]{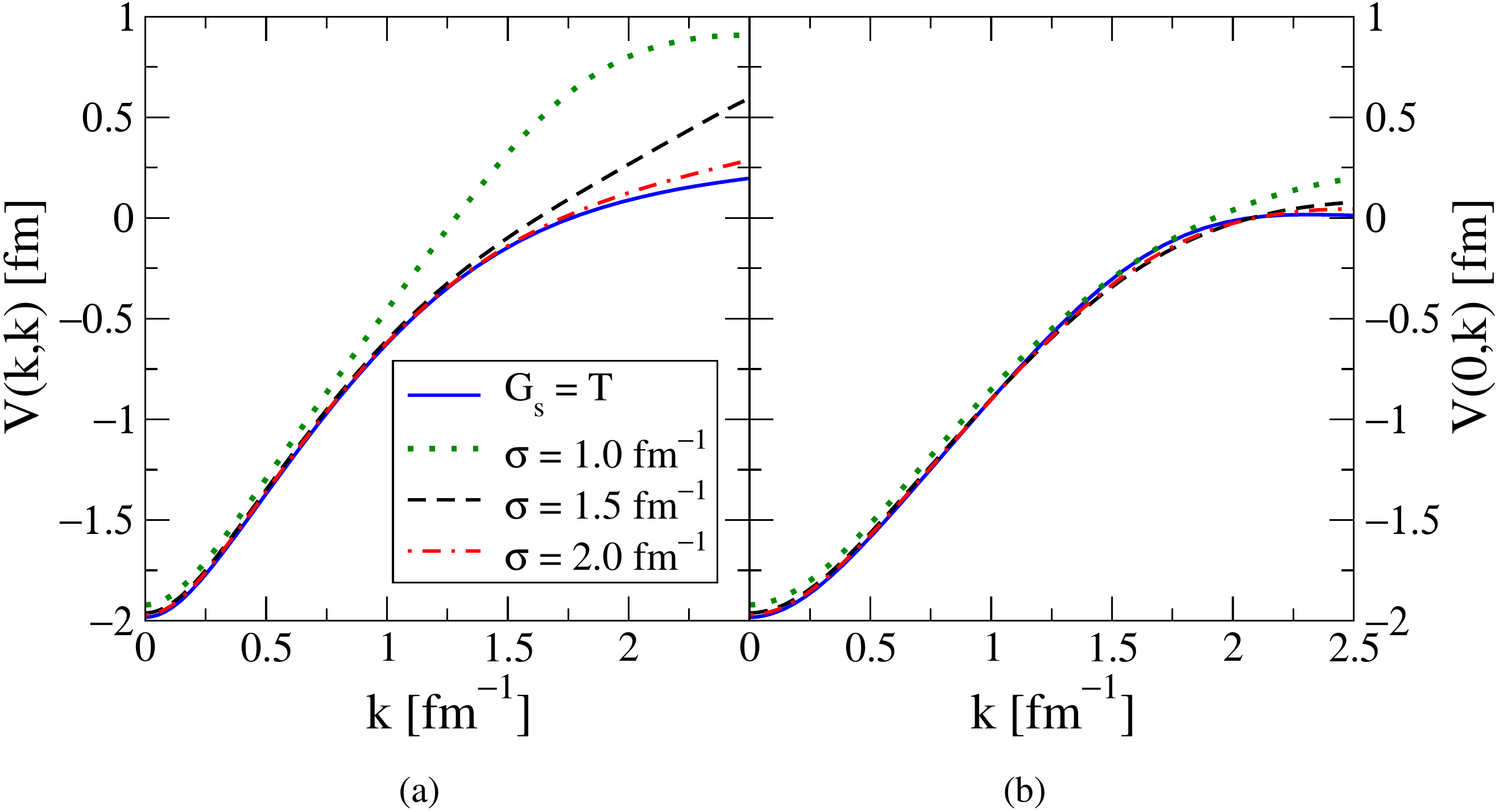}
 \caption{(Color online) Diagonal (a) and off-diagonal (b) momentum-space matrix elements for evolved 
 Argonne $v_{18}$ $^1$S$_0$ potential~\cite{Wiringa:1994wb} with $G_s = T$ and $G_s^{\rm exp}$ with different values of $\sigma$ each evolved  to \(\lambda=2\fmi\). The lines with $\sigma = 3\,\fmi$ are indistinguishable from the ones with $G_s = T$.}
 \label{fig:universalityexp}
\end{figure*}

\begin{figure*}[tbh]
  \subfigure[]{\includegraphics[width=0.26\textwidth]{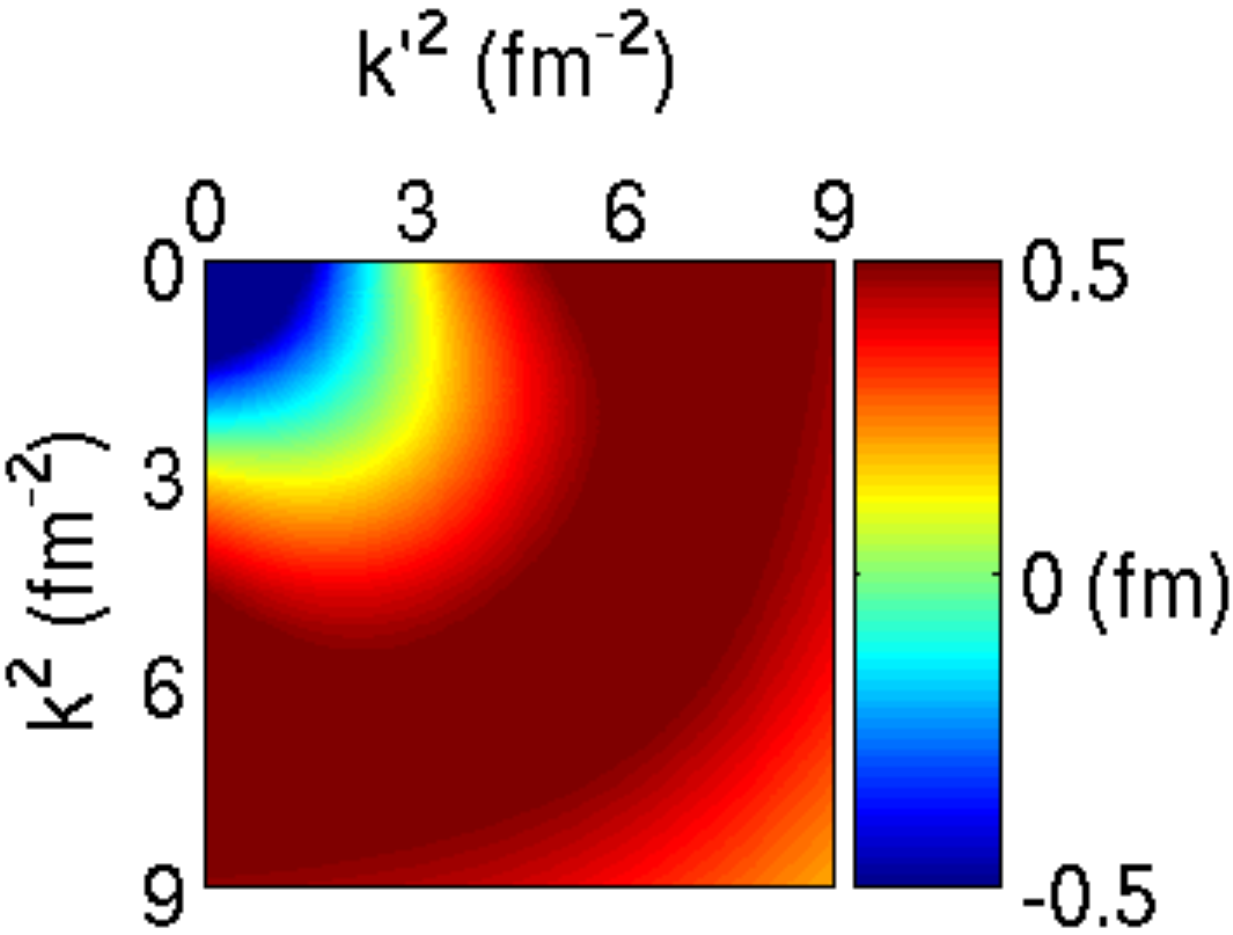}}
  \hspace*{.2in}
  \subfigure[]{\includegraphics[width=0.55\textwidth]{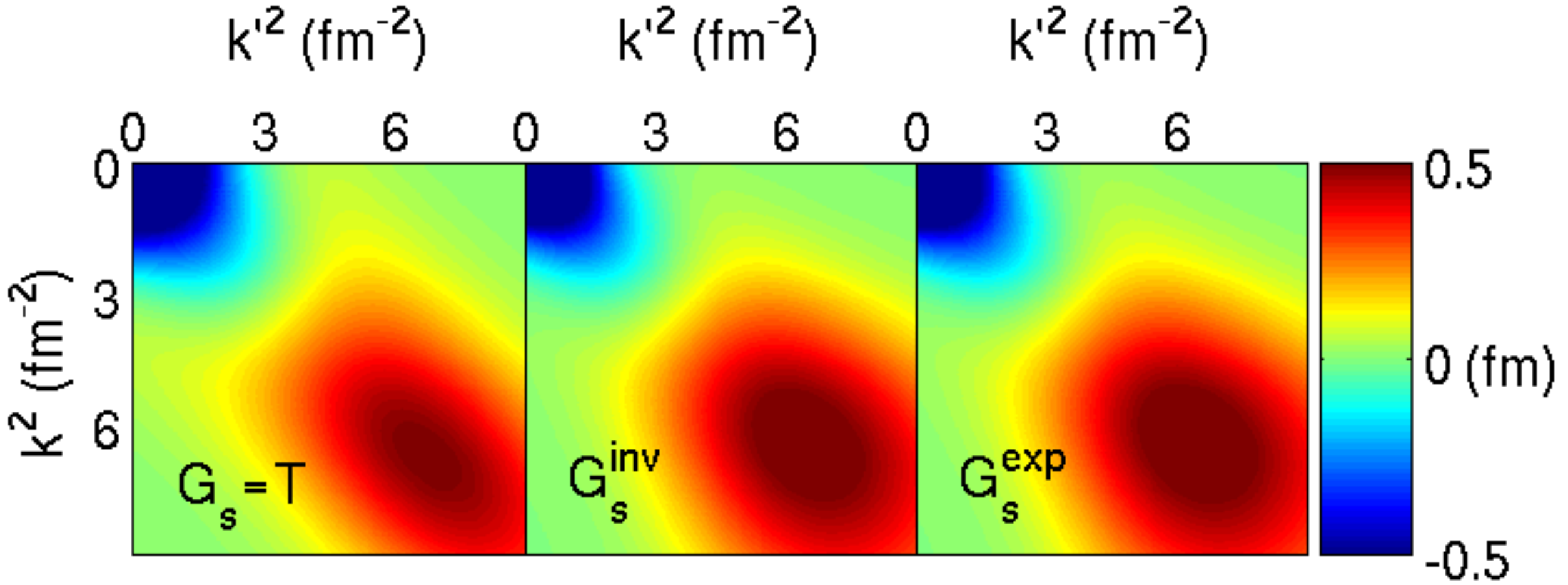}}
 \caption{(Color online) (a) Unevolved N$^3$LO 500 MeV  $^1$S$_0$ potential . (b)  Contour plot showing the evolved N$^3$LO $^1$S$_0$ potential with different generators to \(\lambdaequiv=2\fmi\).  For  $G_s^{\rm exp}$ and $G_s^{\rm inv}$,  \(\sigma=2\fmi.\)\\  }
 \label{fig:evolveN3LO}
\end{figure*}

In Fig.~\ref{fig:evolveAll}
we compare the evolution pattern of the two-body potential in the
$^1$S$_0$ channel with different generators.  Each frame is a representation of the potential matrices in momentum space; where the matrix is zero, there is no coupling between momentum components.   
The initial potential in all cases is Argonne $v_{18}$~\cite{Wiringa:1994wb} and the
value of $\sigma$ is taken to be $2\,\fmi$.  Note that at \(\lambda=4\fmi\), the first matrix plotted here, there is already significant evolution. As the potential is evolved, its high and low momentum components become increasingly decoupled, as expected.
It is evident that the evolved potentials are similar (but not
identical) in the region where $k^2,k'{}^2 < \sigma^2$. When \(\lambda<\sigma\) we find that \(\sigma\) roughly defines the low-momentum region where these generators behave as
$G_{s}=\Trel$.    The minor differences in this low-momentum region can be attributed to the fact that the point in  evolution, \(\lambda\),  needed for these generators to reach the corresponding \(\lambdaequiv\) occurs when  \(\lambda<\lambdaequiv\).  At higher momenta, novel generator evolution is suppressed  relative to $G_{s}=\Trel$.  The patterns here are characteristic of the particular generator and are similar for other potentials and in other channels.

This nature of the evolution is further illustrated by the plots in Figs.~\ref{fig:universalitygenerator} and \ref{fig:universalityexp}, which  show a detailed view of the diagonal and off-diagonal values of the matrices at low momentum for each generator applied to the Argonne $v_{18}$ $^1$S$_0$ potential
. The flow parameter was run to \(\lambda=2\fmi\) with the generators here to distinguish between ambiguities caused by using \(\lambdaequiv\). In Fig.~\ref{fig:universalitygenerator},  the values from different generators agree quite well for $k < \sigma$, where \(\sigma\approx\lambda\).  Moreover, we see that the minor  ``pincushion" effect seen in Fig.~\ref{fig:evolveAll}
  for these generators relative to $G_s = T$  is indeed an artifact of using \(\lambdaequiv\), as each curve falls to zero at approximately the same time.  In  Fig.~\ref{fig:universalityexp}  we focus on \(\GsExp\) versus $G_s = T$ and the fact that they differ for \(\sigma\leq\lambda\).  However, as \(\sigma\rightarrow\lambda\) it is evident that the evolution of the novel generator becomes increasingly similar to  $G_s = T$ in the low-momentum region \(k<\lambda\). For  larger values of \(\sigma,\) these generators become indistinguishable from $G_s = T$.

\begin{figure}[tb]
  \includegraphics[width=\columnwidth]{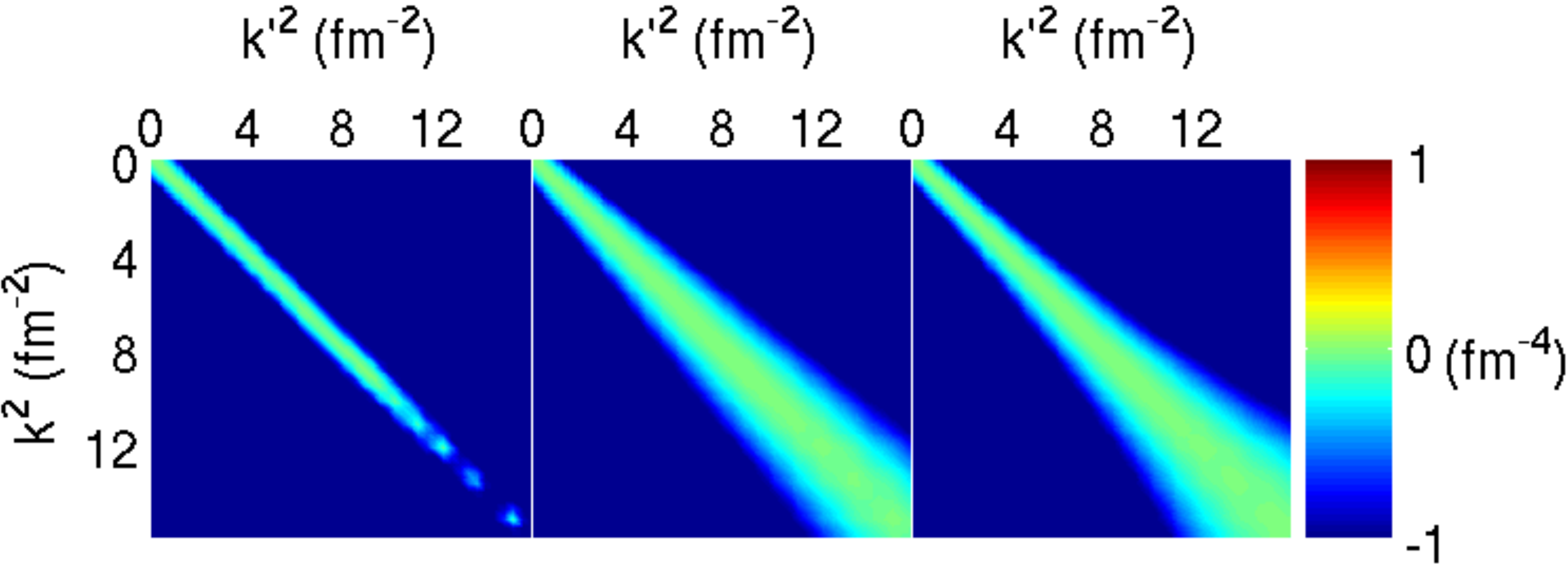}
  \caption{(Color online) Color contour plots of the first term (excluding the factor due to \(V_{s}\)) on the right side of   
  Eq.~\eqref{eq:PWflow}
  for $G_s = \Trel$ (left), $G_s^{\rm inv}$ (middle), and 
   $G_s^{\rm exp}$ (right).
  The last two use $\sigma = 2 \fmi$.}
  \label{fig:first_term}
\end{figure}

\begin{figure}[tb]
  \includegraphics[width=\columnwidth]{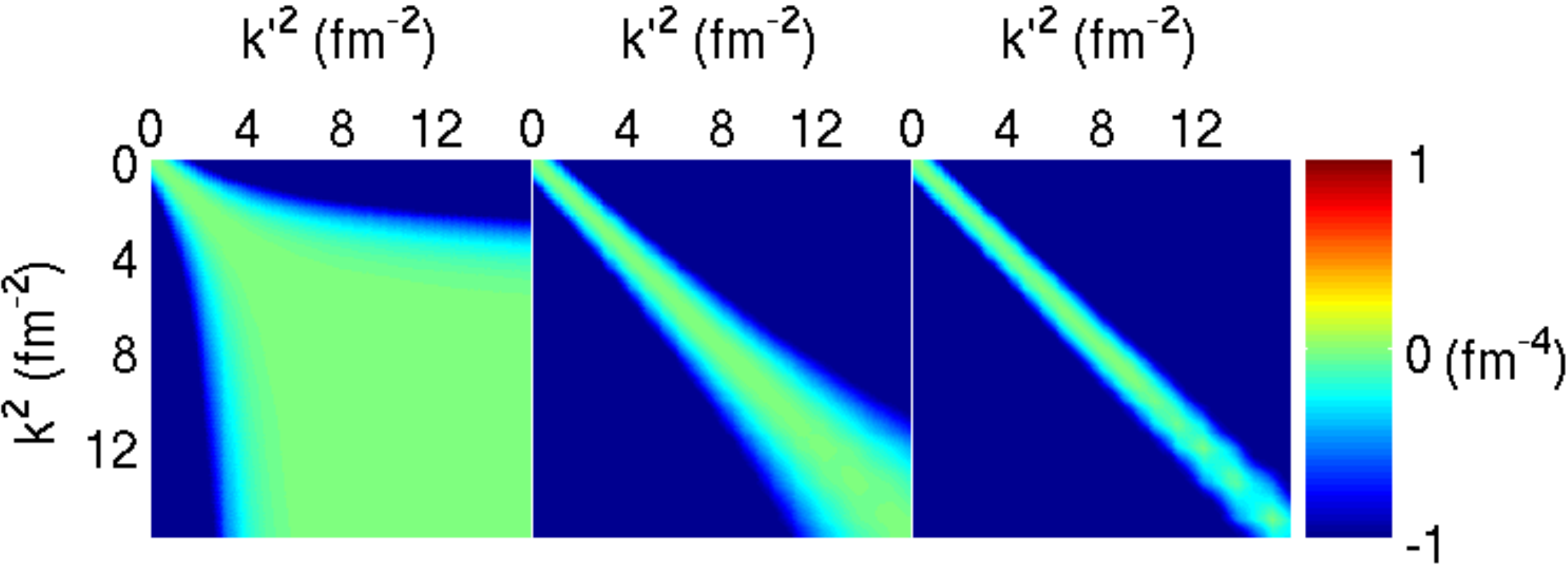}
  \caption{(Color online) Color contour plots of the first terms  
  (excluding the factor due to \(V_{s}\)) on the 
right side of Eq.~\eqref{eq:PWflow} for $G_s^{\rm exp}$ with 
$\sigma = 1, 2,$ and $3 \fmi $ from left to right.}
  \label{fig:Gs_first_term}
\end{figure}

\begin{figure*}[tbh!]
 \includegraphics[width=.8\textwidth]{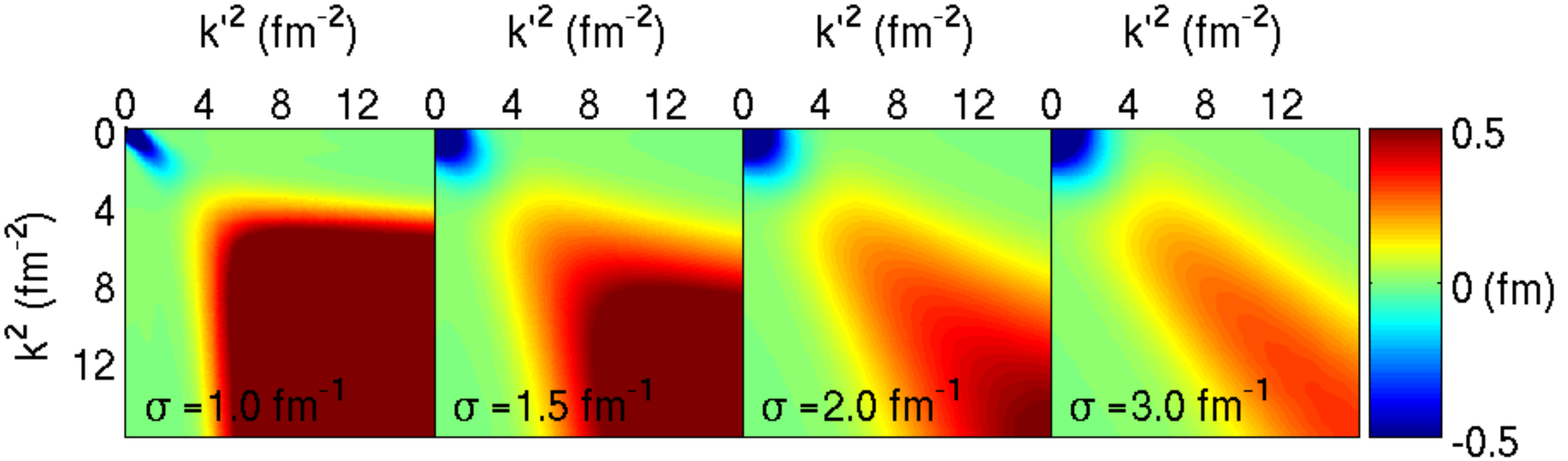}
 \caption{(Color online) Contour plot showing the evolution to  \(\lambdaequiv=2\fmi\) via
  Eq.~\eqref{eq:srgflow}
  starting from the momentum-space Argonne $v_{18}$ $^1$S$_0$ potential~\cite{Wiringa:1994wb}
  using $G_s^{\rm exp}$ from Eq.~\eqref{eq:expgen} with different values of $\sigma$.\\ }
 \label{fig:evolvesigma}
\end{figure*}

\begin{figure}[tb!]
  \includegraphics[width=.9\columnwidth]{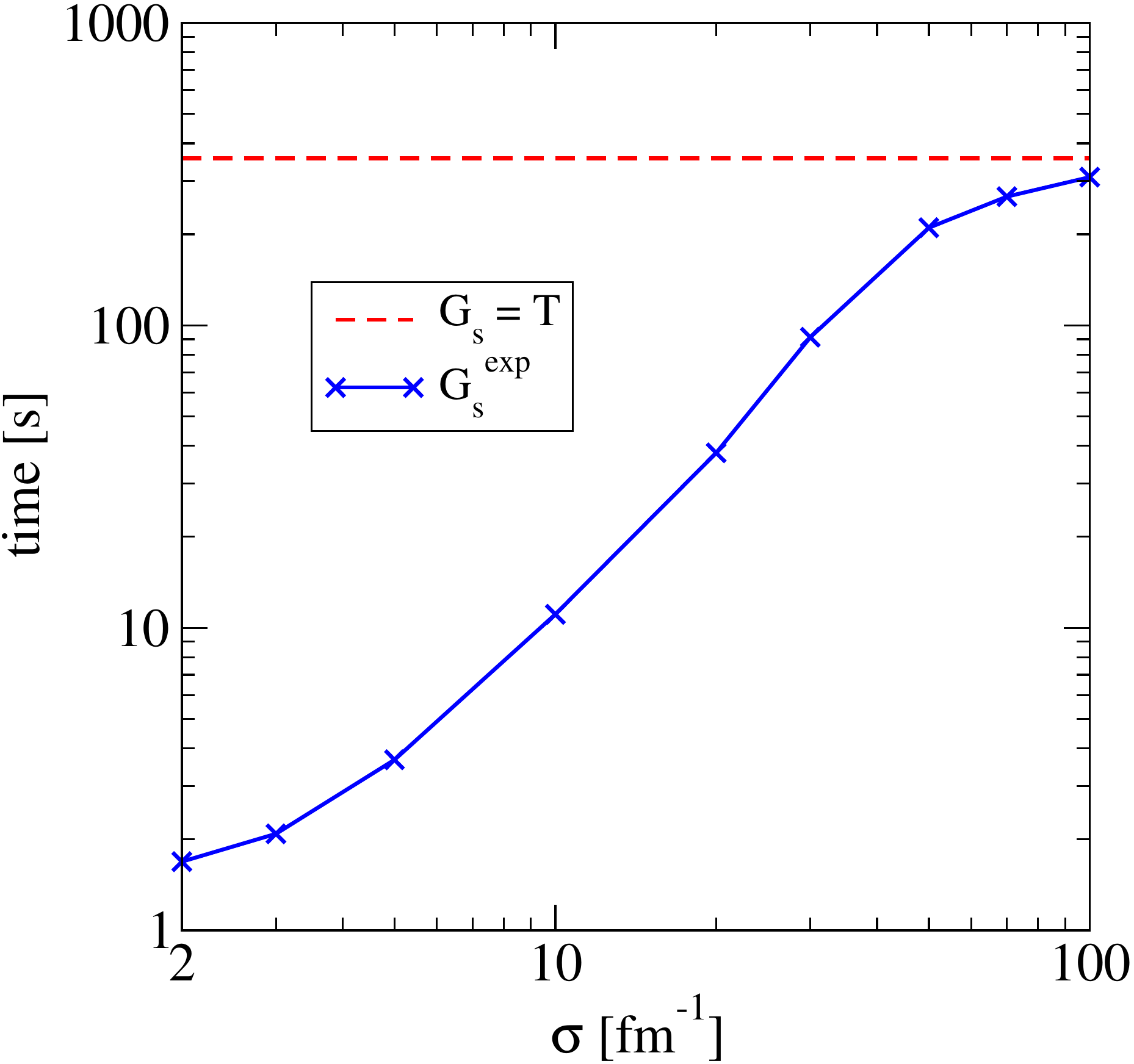}
  \caption{(Color online) Computational time to evolve the Argonne $v_{18}$ $^1$S$_0$ potential~\cite{Wiringa:1994wb}
  with $G_{s}=T$ and \(\GsExp\) to \(\lambdaequiv=2\fmi\)
   as a function of the value of $\sigma$. }
  \label{fig:time2}
\end{figure}

As a demonstration of similar behavior for different NN potentials,  
an evolved potential with different generators 
in the $^1$S$_0$   channel of the N$^3$LO 500 MeV potential~\cite{Entem:2003ft} is shown in Fig.~\ref{fig:evolveN3LO}. Note that the initial potential has significantly less coupling at high momentum compared to 
Argonne $v_{18}$~\cite{Bogner:2009bt}.  As a result, there is correspondingly less improvement in evolution speed. However,  the general features of the evolution patterns with different generators seen
with Argonne $v_{18}$ are also seen for the N$^3$LO potential.

To better understand the evolution process, we need to look further
into the flow equation itself.  Evaluating Eq.~\eqref{eq:srgflow} in a
two-body partial-wave momentum space basis with $G_s = T$ yields
\bea
 \frac{dV_s(k,k')}{ds} &=& - (k^2 - k'{}^2)^2 V_s(k,k')
  \nonumber \\ & & \hspace*{-.6in} \null
   + \frac{2}{\pi} \int_0^\infty q^2 dq\,
    (k^2 + k'{}^2 - 2q^2) V_s(k,q) V_s(q,k')
    \;.
\eea    
In the far off-diagonal region, the first term dominates (this is true for the
ordinary range of $\lambda$ but is modified when $\lambda$ is
comparable with the binding momentum of a bound state).
This implies that each off-diagonal matrix element is driven to zero
as
\beq
  V_s(k,k') \stackrel{k \neq k'}{\longrightarrow} V_{s=0}(k,k')\, e^{-s(k^2-k'{}^2)^2}
  \;.
  \label{eq:driven}
\eeq
For $G_s = f(T)$, these results are modified to
\bea
 \frac{dV_s(k,k')}{ds} &=& - \bigl(k^2 - k'{}^2 \bigr)
   \bigl(f(k^2)-f(k'{}^2) \bigr) V_s(k,k')
  \nonumber \\ & &  \null
   + \frac{2}{\pi} \int_0^\infty q^2 dq\,
    \bigl(f(k^2) + f(k'{}^2 )- 2f(q^2) \bigr) 
  \nonumber \\ & &  \null
    \null \times V_s(k,q) V_s(q,k')
    \; 
    \label{eq:PWflow}
\eea    
and
\beq
  V_s(k,k') \stackrel{k \neq k'}{\longrightarrow} V_{s=0}(k,k')\,
    e^{-s(k^2 - k'{}^2)(f(k^2) - f(k'{}^2))}
    \;.
    \label{eq:drivenf}
\eeq
The difference in the exponents of Eqs.~\eqref{eq:driven} and \eqref{eq:drivenf}
for $k \sim \lambda$ leads to $\lambdaequiv < \lambda$.

The first term of the flow equation (excluding the factor due to the potential) for each of our generators is
shown as a contour plot
in Figs.~\ref{fig:first_term} and \ref{fig:Gs_first_term}.
In Fig.~\ref{fig:first_term} we see that $T$ works uniformly (in $k^2$)
on the entire region of the potential.
With $G_s^{\rm inv}$, there is much less evolution close to the
diagonal.  The plot for $G_s^{\rm exp}$ is similar, but exhibits even less
evolution in the middle region.

It is evident here that the novel generators will result in less evolution at high momenta, as we have seen, and that this
should be a generic result for other partial waves and for higher-body
evolution, because it depends on only kinetic energy differences.
We also see how the value of $\sigma$ controls the degree to which the
operator $G_s$ is similar to $T$.
This is illustrated in Fig.~\ref{fig:Gs_first_term}.
If $\sigma = 1\,\fmi$, only the edges of the potential are
modified; the shape is completely different from $T$.
For $\sigma = 3\,\fmi$, there is the thinnest band on the diagonal,
which is closest to $T$.
At very large $\sigma$, there is a transition to $T$. In the plots of Fig.~\ref{fig:evolvesigma} we see how the final evolved flow is affected by differing choices of \(\sigma\).

The limited evolution at high momenta seen in these generators suggests that the time to evolve to a given decoupling parameter $\lambdaequiv$  should be less for $G_s^{\rm inv}$ or $G_s^{\rm exp}$ than for $G_s = T$.  The dramatic drop in evolution time seen in Fig.~\ref{fig:time1} for the generators makes it apparent that they are more efficient. However, we can also look at how the choice of \(\sigma\) affects their performance.  This  is shown in Fig.~\ref{fig:time2} for $G^{\rm exp}_s$, where we see that the time spent evolving to $\lambdaequiv=2\,\fmi$ decreases as $\sigma$ decreases. As \(\sigma\) becomes smaller, the evolution at high momentum is increasingly limited, which we correlate with improvement in computation time; as \(\sigma\) increases, the evolution time approaches that of \(\Trel\) (as does the flow).  Note that for the Argonne $v_{18}$ potential a large $\sigma$ is needed before $G_s$ is effectively equal to $T$.

In practical applications, one might optimize the trade-off between
decoupling and computational speedup by 
choosing \(\sigma \) to be approximately equal to, or slightly greater than the \(\lambda\) corresponding to \(\lambdaequiv\).  
Then the decoupling properties of \(G_s=\Trel\) are preserved in the low momentum region of interest while still enhancing the computational performance of the evolution by limiting the evolution at high momentum. 
However, this prescription has not yet been tested in detail.

\section{Few-body tests in a one-dimensional model}\label{sec:oneD}

The effects on matrices in a momentum basis demonstrated in
the last section are generic and so 
should carry over to alternative bases and to higher-body forces.
Calculations for realistic three-dimensional few-body systems
are not yet available, but we can test the generators in a 
one-dimensional model of bosons that has proven to accurately predict the evolution of three-dimensional few-body forces~\cite{Jurgenson:2008jp}.  The model we use was originally
introduced in Ref.~\cite{Alexandrou:1988jg} as
a sum of two Gaussians to simulate repulsive
short-range and attractive mid-range nucleon-nucleon two-body
potentials.  It is written in coordinate space as %
\beq
  V^{(2)}(x) = \frac{V_1}{\sigma_1\sqrt{\pi}} e^{-x^2/\sigma_1^2}
    + \frac{V_2}{\sigma_2\sqrt{\pi}} e^{-x^2/\sigma_2^2} 
\eeq
and
in momentum space as
\beq
   V^{(2)}(p,p') = \frac{V_1}{2 \pi\sqrt{2}}e^{-(p-p')^2 \sigma_1^2/8} 
          + \frac{V_2}{2 \pi\sqrt{2}}e^{-(p-p')^2 \sigma_2^2/8} \;,
  \label{eq:gaussians}
\eeq
where $V_1=12$, $\sigma_1=0.2$, $V_2=-12$, and $\sigma_2=0.8$ (see Ref.~\cite{Alexandrou:1988jg} for discussion of units).  Also, in calculations with an 
initial three-body potential, a regulated contact interaction is used.  This is written as 
\beq
   V^{(3)}(p,q,p',q') = c_E\,f_{\Lambda}(p,q)f_{\Lambda}(p',q'),
  \label{eq:gaussians3B}
\eeq
where $c_E$ corresponds to the strength of the interaction and 
\beq
   f_{\Lambda}(p,q) = e^{-((p^2+q^2)/\Lambda^2)^n}.
  \label{eq:gaussians3Breg}
\eeq
The regulator cutoff $\Lambda=2$, and the sharpness of the fall off is set to $n=4$.  We have chosen these parameters for comparison with Ref.~\cite{Jurgenson:2008jp}.

\subsection{Performance}

\begin{table} 
\begin{tabular}{|c|c|c|c|c|}\hline
 \mystrut \textbf{Basis} & \multicolumn{1}{c}{Momentum} \vline& 
 \multicolumn{3}{c}{Oscillator} 
 \vline\\\hline\textbf{System} & $ A=2$ & $ A=2$ & $ A=3$  & $ A=4$ \\\hline
\(\sigma=2\) & 3.3 & 3.3 & 3.8  & 4.1 \\\hline
\(\sigma=3\) &  2.6 &  2.6 &  2.7  & 2.8 \\\hline
\end{tabular}
\caption{  Speed up in model \(A\)-particle one-dimensional (1D) oscillator basis for evolution to $\lambda_{eq}\approx3$, comparing  the ratio of the time to evolve \(G_{s}=\Trel\) versus \(\GsExp\). The results for \(\GsInv\) are very similar.}
\label{table:speedup}
\end{table}

Our first test is to confirm that the enhanced computational performance characteristics of these generators are maintained in the few-body basis.  Table \ref{table:speedup} shows the  speedup obtained using the \(\GsExp \) generator with the model potential described previously.  The performance of the \(\GsInv\) generator is very similar to the \(\GsExp\) results quoted here.
Results are reported at an optimal \(\lambdaequiv\); we do not use a range of \(\lambda\) values here because of the particularities of the oscillator basis convergence properties, which is discussed in more detail in what follows.

\begin{figure*}[tbh]
  \subfigure[]{\includegraphics[width=0.68\columnwidth]{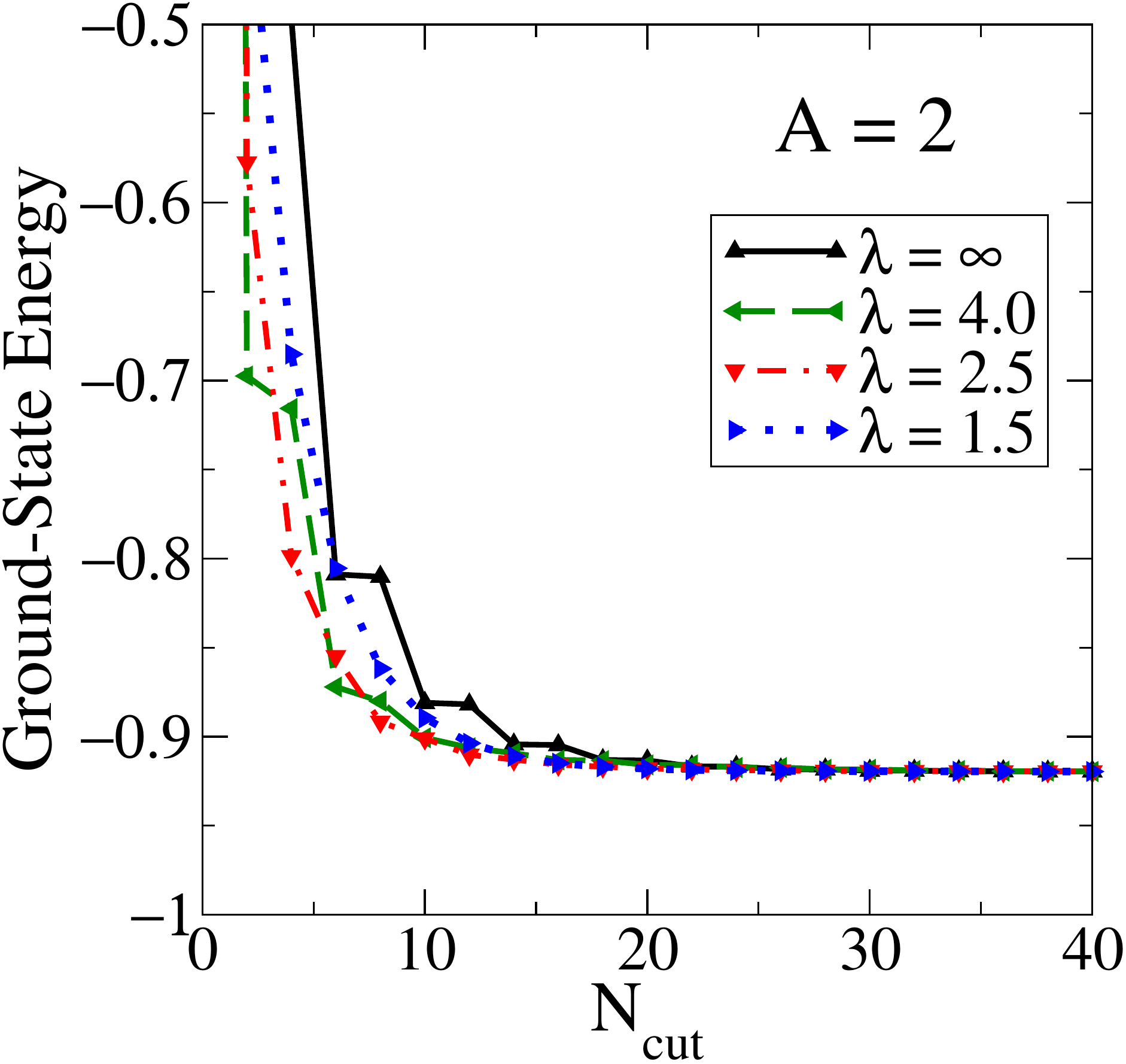}}
  \subfigure[]{\includegraphics[width=0.68\columnwidth]{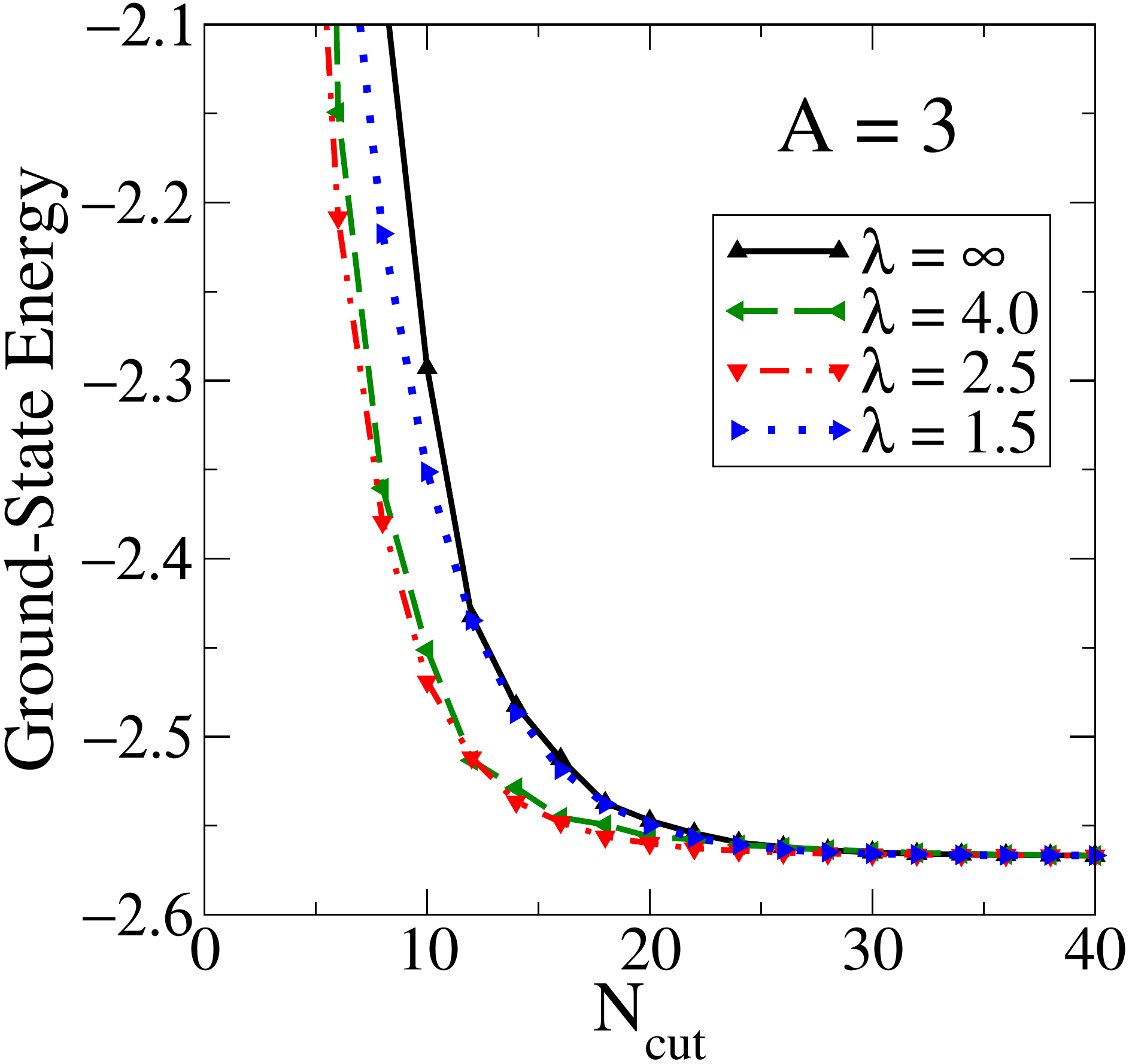}}
  \subfigure[]{\includegraphics[width=0.68\columnwidth]{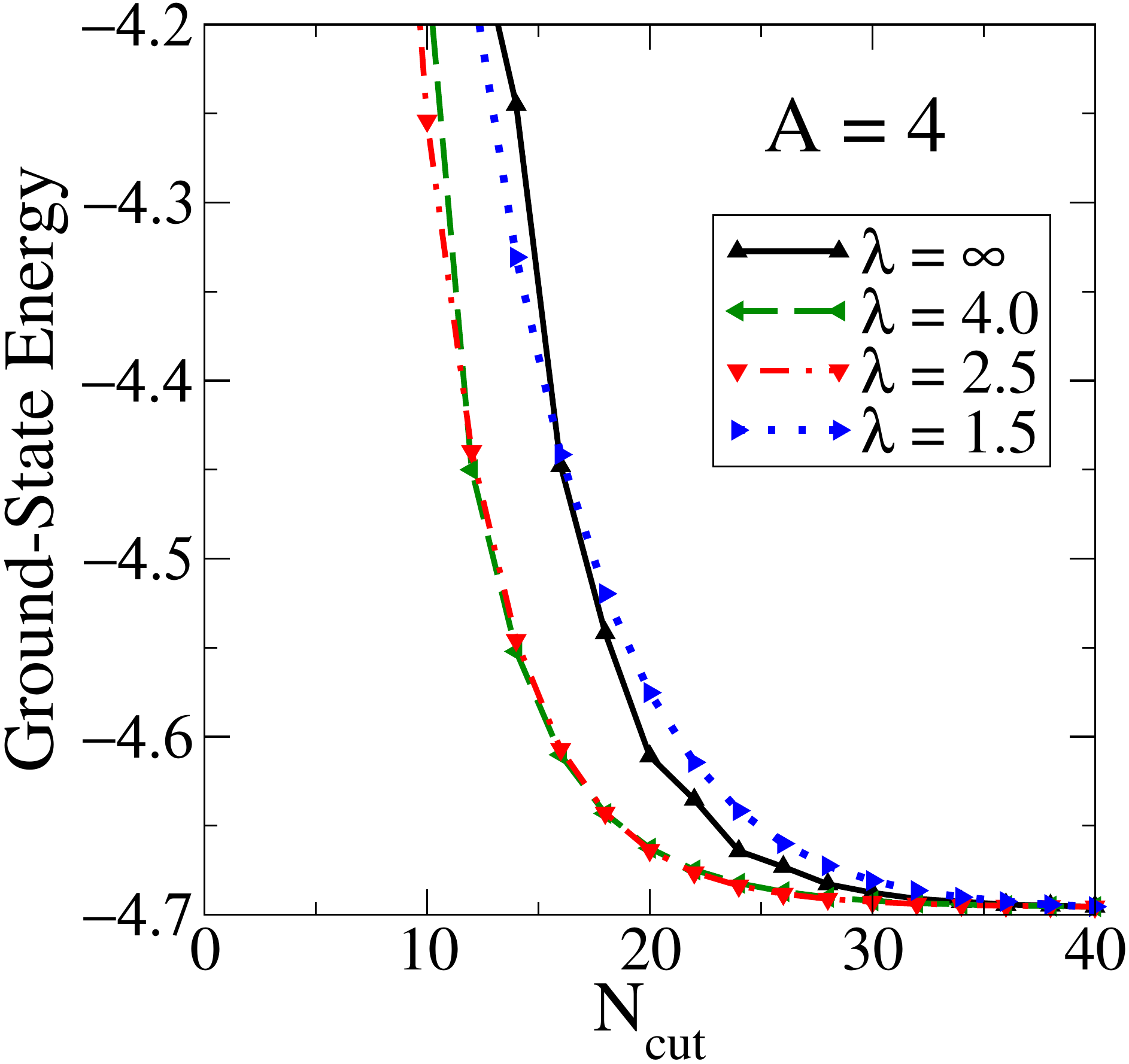}}
  \caption{(Color online) Decoupling using $G_s^{\rm exp}$ with \(\sigma=3\) for (a) $A=2$, (b) $A=3$, and (c) $A=4$. The
   initial two-body-only potential is evolved to each $\lambda$ shown in a basis with \(N_{\rm max} = 40\). Matrix
   elements of the potential are set to zero if one or both states have $N > N_{\rm cut}$ 
   and the resulting Hamiltonian is diagonalized to obtain the ground-state energies plotted.}
  \label{fig:decouplingmanybody}
\end{figure*}

\begin{figure*}[tb!]
  \subfigure[]{\includegraphics[width=0.683\columnwidth]{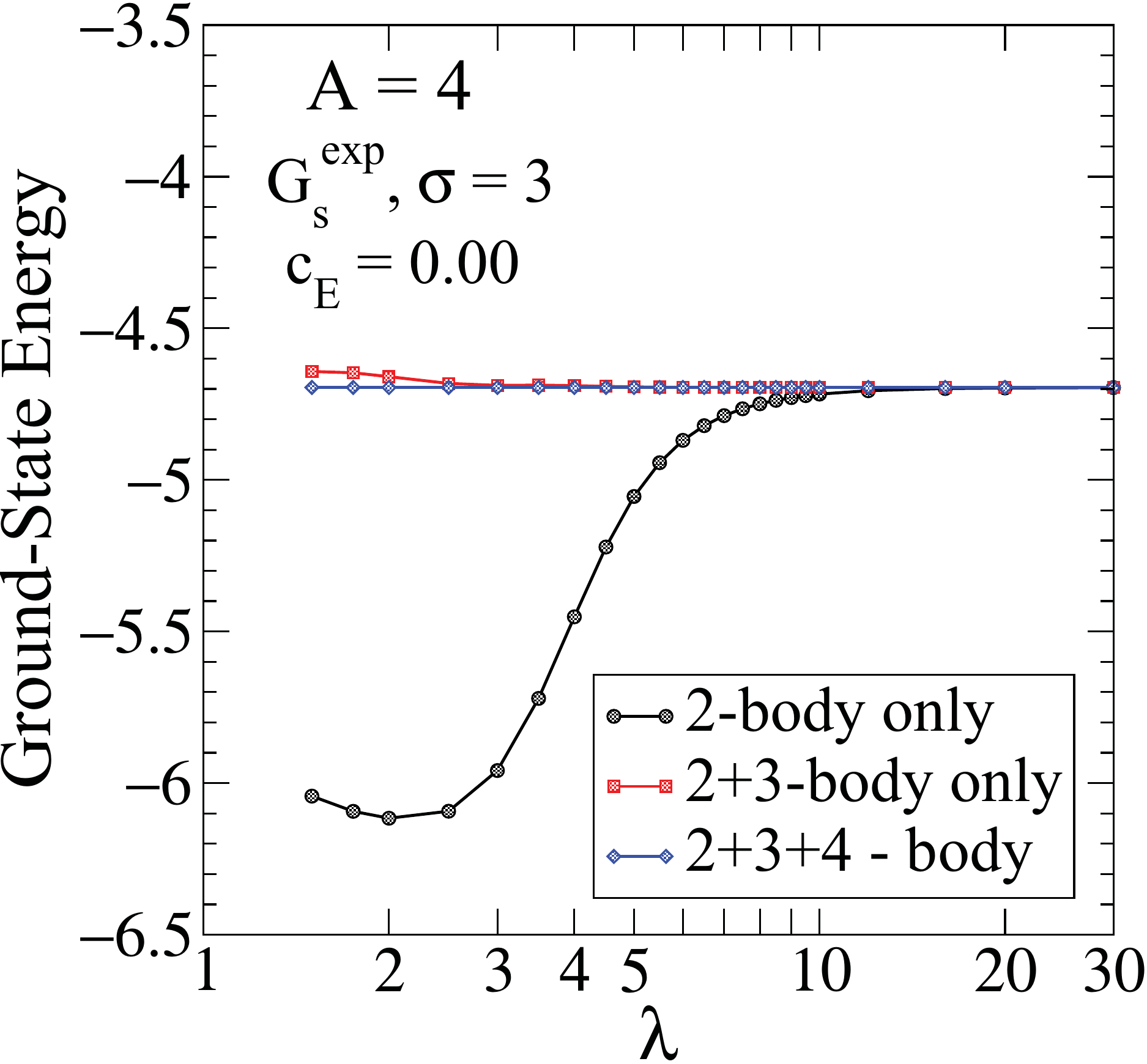}}
  \subfigure[]{\includegraphics[width=0.683\columnwidth]{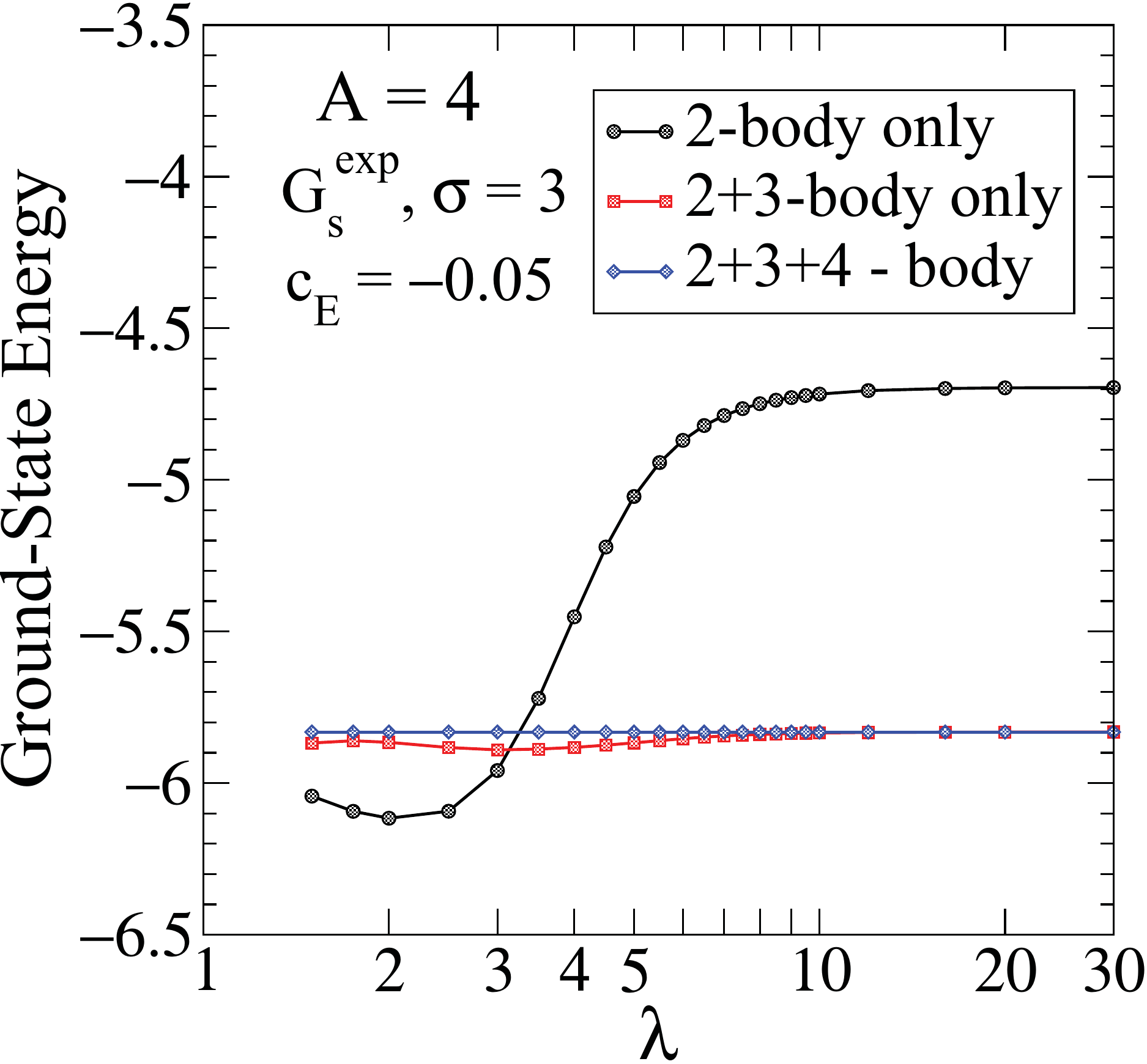}}
  \subfigure[]{\includegraphics[width=0.683\columnwidth]{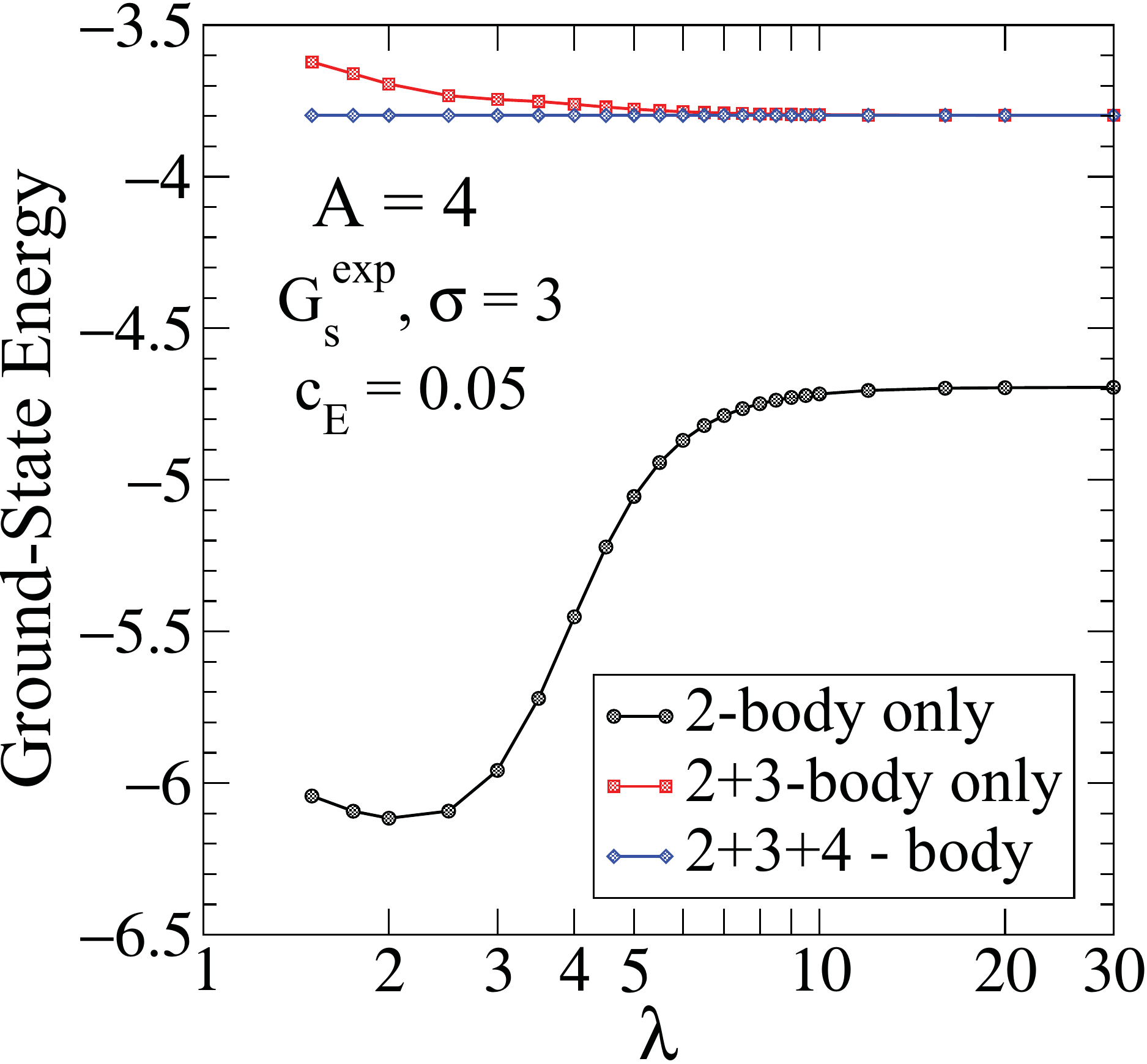}}
  \caption{(Color online) The lowest bound-state energy for a four-particle system as a function of $\lambda$ evolved
  using the SRG $G_s^{\rm exp}$ generator and an $N_{\rm max} = 40$ basis with an initial three-body potential with strength (a) $c_{\rm E}= 0.0$, (b) $c_{\rm E}= -0.05$, and (c) $c_{\rm E}= 0.05$. 
  The (blue) curves with diamonds include the full evolution of the Hamiltonian while the (black) curves with circles use the two-body potential evolved in the two-particle system and the (red) curves with squares use the two- and three-body potentials evolved in the three-particle system. Evolution with $G_s^{\rm inv}$ is almost
  indistinguishable.}
  \label{fig:Ebindwithlambda}
\end{figure*}

We find that the evolution of  the two-body force in two-, three-, and four-particle systems 
with novel generators are all 2.5--4 times faster than the evolution with $G_s = \Trel$ (to the same 
degree of decoupling).  The performance enhancement is relatively basis independent, with the speedup for the momentum and oscillator bases roughly equivalent.  
While the speedup improvement is much smaller than that found for Argonne $v_{18}$,
we do not expect a direct correspondence.
The important point is that the
speedup in the A=2 particle system serves as a good predictor of the speedup in the A=3 and A=4  systems.  Thus, one might expect a similar improvement in few-body oscillator basis calculations with Argonne $v_{18}$ as found previously for Argonne $v_{18}$ in the two-particle partial-wave momentum basis. 
This will be significant as we move to novel generator calculations in  realistic three-dimensional systems.

\subsection{Decoupling}

The measure of performance using these generators depends explicitly on their decoupling properties  in the few-body harmonic oscillator basis relative to  $G_s = \Trel$.  Ultimately, we find the level of decoupling obtained with  $G_s = \Trel$ to be matched by these generators.    

However, the convergence properties of the oscillator basis with respect to SRG evolution, and consequently the issue of selecting an appropriate \(\lambdaequiv\) in this basis, is more complicated than for the momentum basis. The convergence of observables depends on a balance of the ultraviolet (UV) and infrared (IR) cutoffs intrinsic to the choice of a particular oscillator basis.  These cutoffs are given by \cite{Jurgenson:2010wy}   
\begin{equation}
\Lambda_{\rm UV}\sim\sqrt{mN_{\rm max}\hbar \Omega}
\end{equation}
and
\begin{equation}
\Lambda_{\rm IR}\sim\sqrt{\frac{m\hbar \Omega}{N_{\rm max}}}
 \;,
\end{equation}
where \(\Omega\) is the oscillator frequency and \(N_{\rm max}\) is the maximum number of total oscillator excitations in the basis. Thus, a cutoff in oscillator basis states results in two approximate cutoffs in momentum space.  However, the SRG, using the generators considered here, provides only a means to effectively lower the UV cutoff (by decoupling high- and low-momentum degrees of freedom in the Hamiltonian).  As such, convergence is not monotonically improved with respect to evolution in  \(\lambda\). 

 As a measure of the decoupling, we plot the binding energy of the lowest energy state for an \(A\)-particle system with respect to  \(N_{\rm cut}\) for evolutions of the initial Hamiltonian to various  \(\lambda\) (this procedure was carried out for  $G_s = \Trel$ using this model in \cite{Jurgenson:2008jp}).  The actual calculation is carried out by evolving the model interaction to \(\lambda\) in an initial basis large enough so that the binding energy is well converged.  The Hamiltonian is then truncated at \(N_{\rm cut}\)  and the binding energies are calculated in the reduced basis. The value of \(N_{\rm cut}\) refers to the number of oscillator  excitations in the basis and is the oscillator basis equivalent of  the \(\kcut\) parameter in  momentum space, as used in Sec. \ref{sec:NN}. Again, each \(N_{\rm cut}\)  here corresponds to a rough \(\Lambda_{\rm UV}\) and \(\Lambda_{\rm IR}\) truncation in momentum space.   

Results are shown in Fig.~\ref{fig:decouplingmanybody} for the A = 2, A= 3, and A=4 particle systems using 
$G_s^{\rm exp}$ with $\sigma = 3$ for selected values of \(\lambda\).  The signature of decoupling is the improved convergence of the binding energy at smaller \(N_{\rm cut}\) with respect to SRG evolution.  As the interaction is evolved, the degree of decoupling gets better. This is true only up to some value of $\lambda$, however, at which point the degree of decoupling starts to get worse.  It is the latter behavior, introduced by the IR\ cutoff of the oscillator basis, that complicates our efforts to choose a \(\lambdaequiv\) because a one-to-one correspondence with \(\lambda\) is no longer clear.  

Nevertheless, a practical choice can be made by  equating \(\lambda\) with \(\lambdaequiv\) when decoupling is found to be optimal in the \(G_{s}=\Trel\) and novel generator evolutions.
The optimal levels of evolution
happen to coincide at \(\lambda\approx3\) for  \(G_{s}=\Trel\) and these generators.  Thus, the speedup results in Table \ref{table:speedup} were reported at \(\lambdaequiv=3\).  Moreover, given these values for \(\lambda\) and \(\lambdaequiv\), we have chosen \(\sigma=3\) for most of the model space calculations in this section.

 One may note that differences do exist between the \(G_s=\Trel\)  and novel generator decoupling results, particularly at low  \(N_{\rm cut}\)   \cite{Jurgenson:2008jp}. However, the  level at which any of the generators become well converged to the exact
results with respect to \(N_{\rm cut}\) are effectively the same.
Thus, it is reasonable to make the comparisons we have done here to determine \(\lambdaequiv\).

\subsection{Induced Many-Body Forces}

In general, the evolution of an interaction via the SRG leads to induced many-body forces.  This is evident if we examine the second quantized form of the Hamiltonian
\begin{equation}
H=T_{ij}a_{i}^{\dag}a_{j}+V_{ijkl}a_{i}^{\dag}a_{j}^{\dag}a_{l}a_{k}+\cdots
\end{equation}
where the dots indicate that higher-body forces  may also be present in the initial interaction.
When the SRG commutators in Eq.~\eqref{eq:srgflow} are performed, one can see that many-body forces will be induced.
These induced forces could pose a serious problem for few-body calculations because at some point we must truncate the model space in numerical calculations, which will alter the predicted value of observables.  
These can be controlled, however, if there is a hierarchy of many-body forces so that successively larger many-body components are suppressed, and one can include at most one or two  induced pieces to obtain well converged results.  This has been found to be the case for \(G_{s}=\Trel\) and needs to hold for any practical alternative generators.   

A measure of the induced many-body forces (which has been used in previous studies \cite{Jurgenson:2008jp,Jurgenson:2009qs,Jurgenson:2010wy}) is to calculate the ground-state energies of the $A=3$ and A=4 particle systems starting with a two-body interaction and to examine how this energy changes with and without the induced three- and four-body components, as a function of the evolution parameter \(\lambda\). We do this in Fig.~\ref{fig:Ebindwithlambda} with a  plot of the ground-state energy for the four-particle system with the initial two-body interaction embedded and evolved in the $A=2$, A=3, and A=4 body bases with the $G_s^{\rm exp}$ generator.  
The results for $G_s^{\rm inv}$ are virtually indistinguishable.
The curves show that the hierarchy of induced many-body forces is preserved for these generators just as with $G_s = \Trel$ \cite{Jurgenson:2008jp}
(see, however, Ref.~\cite{Roth:2011ar}). This hierarchy also holds for calculations with an initial three-body force and in the $A=3$ particle system.   

In summary, the model results suggest that the advantageous features of SRG evolution with $G_s = \Trel$ can be maintained with the added computational performance of these generators when applied to realistic three-dimensional few-body calculations.

\section{Summary}\label{sec:summary}

In this work, novel generators for the SRG that are functions
of the kinetic energy operator $T$ with an adjustable scale parameter
$\sigma$ were tested.  We found that
functions that reduce to $T$ for basis states with kinetic energy
less than $\sigma$   preserved the good features of $T$, such as decoupling,
but efficiently suppressed evolution for higher
kinetic energies and thereby took much less time to evolve.  
Specific examples were considered,
but other choices with a Taylor expansion starting with $T$ should
give comparable results.
Their action was understood using a simple 
analysis of how the generators directly affect regions of high and
low momentum.  If $\sigma$ is large enough, the generators become equivalent
to $T$.
It is important to note that not only the two-body properties of \(\Trel \) were preserved by these generators, but so were its characteristics in a few-body model space. This includes decoupling and the hierarchy of induced many-body forces, which is critical for applications to larger systems of particles.  

The generators allow us to evolve potentials to much smaller
values of $\lambda$ than previously feasible.
This should enable us to explore 
the transition between {pion\-ful} and pionless regions
of EFT potentials and further test the observations
of Glazek and Perry about evolving past a 
bound state~\cite{Glazek:2008pg}.
The original choice for $G_s$ advocated by Wegner and 
collaborators~\cite{Wegner:1994,Kehrein:2006}
and applied extensively in
condensed matter is the diagonal component of the interaction,
$G_s = \Hdiag(s)$,
\beq
  \langle i | \Hdiag(s) | j \rangle  \equiv
  \begin{cases}
  \langle i | H(s)| j \rangle  & \text{if } i = j \;,\\
    0 & \text{otherwise.}
  \end{cases}
\eeq
In Ref.~\cite{Glazek:2008pg}, it was observed that when evolving a
simple model past a bound state
the Wegner
evolution with $\Hdiag$ will decouple the bound state by leaving it as
a $\delta$ function on the diagonal of the Hamiltonian.  
In contrast, with $G_s = \Trel$ the bound states remained coupled to low momentum  
and were pushed to the lowest momentum part of the matrix.
This behavior was explored in Ref.~\cite{Wendt:2011qj} for leading-order, large-cutoff EFT potentials featuring deeply bound spurious states.
However, it has not been studied for the physical deuteron state,
which requires evolving well below $\lambda = 1\,\fmi$.
This is now easily possible with the replacement of $\Hdiag$
for $T$ in Eqs.~\eqref{eq:invgen} and \eqref{eq:expgen},
although there are as-yet-unsolved
complications from the discretization of
the momentum basis.

The most important next step for the novel generators is to apply
them to evolve realistic few-body potentials, where
speeding up the evolution is desirable because of the
large sizes of the matrices involved.  The generators can be applied
directly to few-particle bases using the method described
in Refs.~\cite{Jurgenson:2008jp,Jurgenson:2009qs,Jurgenson:2010wy}.
Calculations in a one-dimensional model performed here imply that the speed up
carries over to three-body forces and could have a significant impact in making realistic calculations with additional induced many-body forces feasible.

\begin{acknowledgments}
We
thank E. Jurgenson, R. Perry, and K. Wendt for useful comments and discussions.  We also thank E. Jurgenson for the use of his 1D few-body code.  This work was supported in part by the National Science Foundation under Grant Nos.~PHY--0653312 and PHY-1002478, and the UNEDF SciDAC Collaboration under DOE Grant DE-FC02-09ER41586. 

\end{acknowledgments}

%

\end{document}